# A paradigm shift, or a paradigm adjustment? The evolution of the Oleaceae mating system as a small-scale Kuhnian case-study

Alexandre Francq[1], Pierre Saumitou-Laprade[1], Philippe Vernet[1] and Sylvain Billiard[1,2]

1. Univ. Lille, CNRS, UMR 8198 – Evo-Eco-Paleo, F-59000 Lille, France
2. Corresponding author: sylvain.billiard@univ-lille.fr

## Abstract

Kuhn (1962) proposed an evolutionary model to explain how scientific knowledge is built, based on the concept of *paradigm*. Even though Kuhn's model is general, it has been applied only to a few topics in evolutionary biology, especially to broad-based paradigms. We analyze here, through the lens of Kuhn's theory, a small-scale paradigm change that occurred with the resolution of the controversy about the mating system of a Mediterranean shrub *Phillyrea angustifolia* (Oleaceae). We first summarize the different steps of the paradigm change and replace them in the more general context of the sex ratio theory. Second, we show how the different steps of the paradigm changes can be interpreted by Kuhnian concepts and tools. Finally, we discuss the actual and future status of the new paradigm.

## Introduction

Since the publication of *The Structure of Scientific Revolutions* in 1962, Thomas Kuhn's theory of scientific research has been a cornerstone in philosophy of science. Kuhn's model seeks to identify global mechanisms underlying all scientific activities, at all scales, and aims at explaining the efficacy of scientific activity (Kuhn 1970). The Kuhnian model is an evolutionary theory (or a "post-darwinian kantianism", Kuhn 2000) as it is based on historical processes, and because paradigm selection is analogous to natural selection. Kuhn's theory is also based on a realistic approach that humanizes scientific research, thus making it more plastic and less idealized. Even though the general applicability of Kuhn's theory was debated very early after the publication of *Structure* (see Shapere 1964 for a critical statement about *incommensurability,* one of the fundamental concept of the theory), Kuhn's model is certainly one of the most successful epistemological frameworks, especially among scientists themselves.

Biology is strikingly absent from *Structure*, certainly because Kuhn was known to be more comfortable with the history of physics and chemistry. For more than fifty years, researchers have been trying to apply the Kuhnian model to different Biology fields, especially population biology, at different scales (Fig. 1, see also the *epistemological meta-paradigms* in Friedman, 2002 for an alternative organization). The adoption of the Darwinian and Neo-Darwinian theory, high level paradigms (Level 2, Figure 1), received much attention (*e.g.* Jacob 1976; Greene 1981, Gayon 1992; Jablonka and Lamb 2005; Morange 2017, Ruse 1970, 2018; Tanghe et al. 2021). Ruse (2018) in particular argued that Darwinism should certainly not be considered as a paradigm but rather as an example of scientific consilience. Bertoldi (2018) further proposed that the Kuhnian category of paradigm does not allow us to accurately portray the specificities of Darwin's theory.

Besides the study of Darwinism, Kuhnian analyses of other population biology issues are scarce: the issue of the unit of selection in evolution (level 3 in Fig. 1; see Lloyd 2020 for a review; see Ruse 1987 for the particular case of sociobiology); the pace of evolution and the theory of the punctuated equilibria (Ruse 1989); Hoquet (2020) partly analyzed the status of Bateman (1948)'s principle (a Level 3 paradigm) especially focusing on the social components of the associated paradigm-based research; Finally, Avise (2014) collected, reviewed and evaluated the importance of a large number of paradigm changes in evolutionary genetics, from different levels (Levels 2 to 4, Figure 1). However, its choices and evaluations of paradigm changes were mostly subjective, as confessed by the author himself. He did not precisely use the Kuhnian model to determine what could be considered as a paradigm, whether a scientific change should be considered as a paradigm shift, or how to evaluate the importance of a shift.

Our goal in this paper is to provide a new Kuhnian analysis of a low level paradigm from population biology: the evolution of mating systems in plants applied to Oleaceae species, a paradigm embedded in the sex ratio theory (Fisher 1930), one of the most celebrated evolutionary theory (Edwards 1998). This paradigm has original specificities which makes it an original and perfect candidate for an analysis of the evolution of a scientific paradigm in evolutionary biology: it is a small scale and low level paradigm (Level 4, Figure 1); it is local in the sense that the involved scientific community and the associated literature is limited; it is actual as a crisis of this paradigm was only recently resolved (although the crisis is still marginally ongoing).

This paradigm crisis followed directly the basic idea of the sex ratio theory put forward by Fisher (1930): producing an offspring in sexual species necessarily involves the fusion of both a male and a female gametes. As a consequence, the sex ratio in a population should be balanced 1:1. Indeed, if one sex is rarer than the other, an individual which would give more offspring of the former sex would have more grandchildren and would thus be favored by natural selection. The mating and sexual systems of the shrub *Phillyrea angustifolia* (Oleaceae) challenged the paradigm of plant mating system evolution as a subcase of the sex ratio theory. The controversy, which will be detailed below, was centred on the observed frequency of males in *P. angustifolia* natural populations which was much higher than expected (Pannell 2002): males were as frequent as hermaphrodites, which was not possible under the sex ratio theory in androdioecious populations (a population is androdioecious when hermaphrodites and male individuals co-occur). Hence, either the observations were wrong (*i.e.* hermaphrodites were in fact functionally females, which would give a 1:1 sex ratio, in accordance with theoretical predictions), or the quantitative theoretical predictions from the sex ratio theory were not correct because not adapted to the case of *P. angustifolia*. The controversy was resolved after the discovery of the link between sex determination and a diallelic self-incompatibility system in this species (Saumitou-Laprade et al. 2010). From an epistemological point of view, this raises the question whether this resolution constitutes a simple paradigm adjustment or a paradigm shift. Pannell and Korbecka (2010) claimed, without a proper analysis of the criteria proposed by Kuhn and others (*e.g.* Wray 2021), that only a paradigm adjustment was necessary (*i.e.* "normal science" is going on).

In this paper, we first detail the controversy about the mating system in *P. angustifolia*, especially highlighting the different crucial steps of the paradigm changes (Figure 2). Second, we study and analyze this controversy through the lens of Kuhn's theory. Third, we discuss the actual and future status of the paradigm changes regarding the case of *P. angustifolia*. In particular, we discuss whether the discovery of the diallelic self-incompatibility system (DSI) in *P. angustifolia* constitutes an adjustment of the previous paradigm, as suggested by Pannell and Korbecka (2010), or a paradigm shift. We also describe current and possible future resistances regarding research on the Oleaceae mating system evolution based on our understanding of Kuhn's theory. Overall, a Kuhnian analysis of this small-scale case-study offers a unique opportunity to analyze how science works in action, to study some phenomena that are rarely observed for high level paradigms (*e.g.* scientists' conversion from the old to the new paradigm), and to thoroughly analyze the roles played by the confrontation between models and data in a paradigm shift. It also raises the question whether the Kuhnian model applies to any paradigm levels, or how important should a scientific change be to be considered as a paradigm shift.

## I. *The* Phillyrea angustifolia *controversy, its resolution and extensions*

**Normal science: sex ratio theory and plant mating evolution.** Understanding the evolutionary mechanisms explaining why sex ratio should be balanced between males and females in dioecious species has a long history that dates back to Darwin (see Edwards 1998 for a review and references therein). Fisher (1930) stated that the mechanism is simple: since producing offspring by sexual reproduction necessarily needs a male and a female gametes, if the sex ratio is unbalanced, then one individual of the rare sex would have more offspring, on average, than an individual of the other sex. Hence a mutation that would bias the sex ratio at birth in favor of the rare sex would be advantaged by natural selection, because individuals bearing such a mutation will have more grandchildren on average. The evolutionary stable equilibrium sex ratio is thus expected to be 1:1.

Departures from the 1:1 prediction are observed in many different species. As a consequence, Fisher (1930)'s paradigm suffered many changes to explain unbalanced sex ratio. For instance, it has been necessary to distinguish between sex ratio at birth (*i.e.* the ratio of males *vs*. females in offspring) and operational sex ratio (the ratio of competing males *vs*. females which can be fertilized) to explain why sex ratio is slighty male-biased in human populations (such a bias being compensated by a higher male death rate before sexual maturity, *e.g.* Ritchie and Roser 2019), or why sex ratio can be highly skewed when a population is structured (Hamilton 1967). Actually, sex ratio theory is regarded more generally as a sex allocation theory (Charnov 1982), *i.e.* the ratio of resources allocated to males *vs*. females gametes by individuals taking into account life history traits, environmental conditions, and interactions between individuals.

In the sex ratio theory, plant mating systems evolution received a particular attention, mostly because most plant species show hermaphroditic individuals. Hermaphroditism was a particular issue for the sex ratio paradigm because it makes possible 1) self-fertilization, and its counterpart strong inbreeding depression, which respectively affect male fitness, because less ovules are available for siring, and female fitness since inbred offspring are less fit; 2) differential resource allocation between male and female gametes within individuals, which can lead to highly skewed pollen-ovule ratio or male *vs*. female reproductive organs production; 3) A generalized definition of sex ratio as the relative frequency of a pair (or a triplet) of genders within a population: males, females and/or hermaphrodites.

After the seminal work by Lewis (1941), new questions raised in the 70's-early 80's about plant mating evolution considering those three consequences of hermaphroditism within the sex ratio theory, especially regarding the expected frequency of unisexual individuals (males or females) when they co-occur with hermaphrodites in natural populations (populations with hermaphrodites and males, hermaphrodites and females, hermaphrodites and both males and females are respectively called gynodioecious, androdioecious, and trioecious). Among others, Lewis (1941), Lloyd (1975), Charnov et al. (1976) and Charlesworth and Charlesworth (1981) provided theoretical quantitative predictions about the expected frequencies of unisexual individuals in populations, and the conditions under which unisexuals are expected to be maintained with hermaphrodites. In other words: what is the evolutionarily stable sex ratio in androdioecious and gynodioecious populations?

As expected by quantitative models and confirmed by empirical observations, gynodioecy and androdioecy are not symmetrical mating systems (reviewed in Pannell 2002, Dufay and Billard 2011). From a theoretical point of view, androdioecy is expected to be less stable (and thus rarer) than gynodioecy in plants. In addition, males are expected to be rare in androdioecious species, *i.e*. the sex ratio is expected to be highly skewed towards hermaphrodites. This is due to three main factors. First, because of the Bateman's principle that posits that males' fitness is mostly limited by mating opportunities, whereas females' fitness is mostly limited by resource acquisition (Bateman 1948). Second, since males can only have offspring by their male gametes, while hermaphrodites can give offspring by both their male and female gametes, males should sire at least twice as many offspring as hermaphrodites to be maintained in a population (Lloyd 1975). This means that resources allocation to male gametes production (or efficiency) should be largely higher in males than in hermaphrodites. Third, because hermaphrodites can self-fertilize, it gives even less opportunities for males to sire available ovules, and it is easier for hermaphrodites to produce offspring through both their male and female gametes. Hence, as self-fertilization increases, it is expected that males are rarer and that androdioecy is less stable (Charlesworth and Charlesworth 1981).

As reviewed by Pannell (2002), empirical observations totally agree with theoretical predictions in flowering plants: androdioecious species are rare, males show low frequencies in natural androdioecious populations, and males show an obvious larger production of male gametes than hermaphrodites. Overall, it means that the sex ratio theory applied to plant mating system perfectly fits observations. There was however one exception that challenged the theory for two decades, as acknowledged by Pannell (2002): the Mediterranean shrub *Phillyrea angustifolia* (Oleaceae).

**Crisis: Sex ratio paradigm's predictions applied to *P. angustifolia* populations.** In a nutshell, the sex ratio paradigm predicts that:

1) In dioecious populations, the male *vs*. female sex ratio should be balanced 1:1, unless

under particular situations;

2) In androdioecious populations, the male *vs*. hermaphrodite sex ratio should be highly skewed towards hermaphrodites, unless males show a much higher production (or efficiency) of male gametes relatively to hermaphrodites.

Nothing fitted in *P. angustifolia*. Gerber and Kieffer (1896) and Lepart and Dommée (1992) showed that three southern France natural populations of this species were morphologically androdioecious, since they were composed of individuals showing flowers with only male organs (males) or with both male and female organs (hermaphrodites). They also showed that in these three populations hermaphrodites were male fertile when outcrossed onto another hermaphrodite. Unexpectedly for an androdioecious species (Pannell 2002), 45 populations from Southern France, Portugal and Spain showed that the distribution of the sex ratio within this species ranged between 0.34 to 0.77, 0.49 on average (Strasberg 1988; Lepart and Dommée 1992; Pannell and Ojeda 2000; compiled in Husse et al. 2013). Even more surprising were the estimates of the male advantage in their production (or efficiency) of male gametes relatively to hermaphrodites: it was estimated to lie between 1 and 2 by Vassiliadis et al. (2000a) and Pannell and Korbecka (2010), a value largely incompatible with the observed sex-ratio and the expectations from the sex-ratio paradigm (Lloyd 1975). This generated a paradigm crisis.

From there, two research programs followed two different paths. A first research program hypothesized that, following the sex-ratio paradigm, since the sex ratio in *P. angustifolia* is approximately balanced, then this species should be dioecious and not androdioecious: hermaphrodites *morphologically* show both male and female reproductive organs, but they should be *functionally* females (Pannell 2002), a situation referred as *cryptic dioecy*, in contradiction to Lepart and Dommée (1992)'s conclusions. Roughly speaking, posing that the paradigm and its quantitative predictions are true, then observations should be wrong. Considering hermaphrodites were in fact females was the least costly explanation for maintaining the established paradigm. At the same time, using genetic markers and paternity analyses, Vassiliadis et al. (2002) assigned both parents based on genetic similarities between adult trees and seedlings. They detected that a substantial amount of seeds were effectively sired by hermaphrodites, and thus concluded that hermaphrodites were *functionally* and not only *morphologically* hermaphrodites, in agreement with Lepart and Dommée (1992).

Yet, doubts remained. Verdú et al. (2006) performed similar paternity analyses on another supposedly androdioecious species with a balanced sex ratio, hence with the same paradoxical observations than *P. angutifolia* (the ash tree *Fraxinus ornus,* also an Oleaceae). Even though Verdú et al. (2006) found that the proportion of seeds sired by hermaphrodites was substantial, they concluded that "the 1:1 sex ratios of *F. ornus* populations indicate that the species is cryptically dioecious". They further speculated that, since hermaphrodites were functional, a high *ovule discounting* was necessary, *i.e.* a very high mortality rate of seeds or offspring produced by a cross between hermaphrodites. Assuming an ovule discounting is costly because there was no evidence of such a mechanism actually occurring, and because it would imply a very substantial mating cost, barely compatible with Darwinian selection, a higher level paradigm (Figure 1). By extrapolation, they also concluded that, most likely, a similar mechanism should occur in *P. angustifolia*, and therefore that it also certainly was cryptically dioecious.

In another paper, Pannell and Ojeda (2000) proposed to explain the paradox of *P. angustifolia* by a phenomenon that could increase male advantage: males would tend to flower more often than hermaphrodites, and consequently would increase their opportunity to sire ovules produced by hermaphrodites. A higher flowering rates in males would also bias the estimation of the sex-ratio towards males, suggesting that methodological limits could play a role in the paradox. However, Pannell and Ojeda (200) did not provide quantitative predictions for the effect of differences in flowering rates on observed sex ratio. To what extent it might explain high males frequency in *P. angustifolia* populations was thus unclear. Overall, cryptic dioecy was the least costly explanation that makes the paradigm's predictions compatible with data.

A second research program took another path and challenged the sex ratio paradigm itself, especially the bases of Lloyd (1975)'s model. Fundamentally, the question remained: what factor could rebalance the male advantage between males and hermaphrodites 1) without making hermaphrodites cryptically females, and 2) such that the production (or efficiency) of male gametes by males are similar to that of hermaphrodites?

Vassiliadis et al. (2000a, 2002) observed that the siring success was homogeneous among male individuals, whereas it was highly variable among hermaphrodites. This observation led Vassiliadis et al. (2000b, 2002) to hypothesize that self-incompatibility could be a mechanism involved in androdioecy in *P. angustifolia*. Self-incompatibility (SI) is a mechanism very common in flowering plants which prevents mating between individuals belonging to the same self-incompatibility group. SI is a self-/non-self recognition system which makes mating possible between compatible genotypes only: individuals can cross only if they do not share the same self-incompatibility phenotype, in other words if they do not belong to the same SI group (*e.g.* individuals from the $G_a$ self-incompatibility group can only sire and be sired by individuals from the $G_b$ SI group). A given individual expressing SI is thus necessarily unable to self-fertilize.

The idea of Vassiliadis et al. (2000b, 2002) relies on the following mechanism: if hermaphrodites express SI groups, while males do not or express a SI group unique to them, then males automatically have a fertilization advantage relative to hermaphrodites. Indeed, since males could sire all hermaphrodites while hermaphrodites can only sire a part of all hermaphrodites in the population, pollen emitted by males would have a higher chance to successfully fertilize an ovule than pollen emitted by hermaphrodites. In addition, SI prevents self-fertilization which precludes a potential extra advantage for hermaphrodites (Charlesworth and Charlesworth 1981).

Vassiliadis et al. (2000b) showed in a model that their hypothesis could partly solve the paradox: males could indeed reach high frequency if the number of SI groups expressed by hermaphrodites was small. However, some information was still missing. First, there was no direct evidence of SI in *P. angustifolia*. Second, even if SI indeed exists in *P. angustifolia*, the number of SI groups in the hermaphrodites should be determined. And here another difficulty arises: in plant mating system literature, SI plant species typically show several dozen SI groups (Castric and Vekemans 2004), which would make Vassiliadis et al. (2000b)'s predictions incompatible with data. Hence, the picture was still incomplete, and the paradox remained unsolved.

**Solving the paradox: Diallelic self-incompatibility and segregation distortion.** The paradox was resolved in two steps. The first step was the discovery of a diallelic self-incompatibility system (DSI) with only two homomorphic SI groups of hermaphrodites ($G_a$ and $G_b$), and no SI in males (Saumitou-Laprade et al. 2010). Such an homomorphic DSI system was unexpected for two reasons. First because only heteromorphic SI systems with two SI groups were known in flowering plants (distyly or heterostyly). Second, because all known homomorphic SI in flowering are multiallelic. The discovery of the DSI was also surprising because, with a single simple mechanism, males automatically compensate their fitness disadvantage compared to hermaphrodites: hermaphrodites can reproduce through their ovules and pollen, while males can only reproduce through their pollen, but hermaphrodites can only sire approximately half of the hermaphrodites (the ones of the other SI group), while males can sire all hermaphrodites. Pannell and Korbecka (2010) and Husse et al. (2013) introduced the DSI into the Lloyd (1975)'s model (or an equivalent population genetics model). They showed that indeed, DSI increases the expected frequency of males in populations. However, once again, not to the extent observed in natural populations.

The second step consisted in the observation of sex ratio distortion at birth, which was less surprising than the DSI because segregation distortion was already observed multiple times in angiosperms. More precisely, it was observed that the inheritance of sexual phenotypes from parents to offspring did not follow Mendelian segregation rules for one particular crossing (Billiard et al. 2015): offspring from the mating between males and hermaphrodites from group $G_b$ are all males. All other crosses with hermaphrodites from group $G_a$ produced progeny with a proportion of hermaphrodites $G_a$, $G_b$, and/or males as expected under Mendelian segregation where two unlinked loci controlled for DSI and sex determination. Billiard et al. (2015) introduced this distortion segregation into Husse et al. (2013)'s model. They showed that the combination of the DSI and the distortion segregation gives expected males frequency compatible with observations in natural populations. This finally solved the paradox from the observation of a balanced sex ratio 1:1 in the androdioecious *P. angustifolia*.

Thanks to a combination of experiments in controlled conditions, theoretical quantitative modeling, and observations in natural populations, the research program that resolved the paradox in *P. angustifolia* allowed the discovery of a brand new biological phenomenon: the homomorphic DSI in flowering plants. Analogous mating systems were already known in fungi, ciliates, yeasts or green algae, but not in angiosperms (Billiard et al. 2011, 2012). This research program is an illustration of the scientific fecundity of confronting data with theory. This research program also allowed the discovery of a clear association between two mechanisms controlling matings which are common in angiosperms, but generally considered separately: sexual (males, females or hermaphrodites) and self-incompatibility. Finally, this research program was pursued by the study of mating and sexual systems in other Oleaceae species. In all studied species, the existence of a DSI system was demonstrated either at the prezygotic stage by controlled stigma test, or at the postzygotic stage by paternity analysis of seeds produced by controlled crosses and/or open pollination: the androdioecious manna ash *Fraxinus ornus,* which also shows high male frequencies, (Vernet et al. 2016); the hermaphroditic olive tree *Olea europaea* (Saumitou-Laprade et al. 2017a; Besnard et al. 2020; Mariotti et al. 2021); the morphologically polygamous but functionnally dioecious common ash *Fraxinus excelsior* (Saumitou-Laprade et al. 2018); and the hermaphroditic privet *Ligustrum vulgare* (De Cauwer et al. 2021). No other case of distortion segregation was yet detected. Hence, *P. angustifolia* is on the one hand still a particular case because of this association between sex ratio distortion and DSI, but on the other hand, it is representative of the Oleaceae family since it shares the DSI with all Oleaceae species studied so far.

**Reorganization of the scientific activities: what did the resolution of the *P. angustifolia* paradox change?** Solving the paradox needed two paradigm changes. First, incorporating that *P. angustifolia* population was structured both by sexes (males and hermaphrodites) and by SI groups for hermaphrodites ($G_a$ and $G_b$), and that mating relationships were asymmetrical (hermaphrodites can only sire one of the two SI groups, while males can sire both SI groups). Whether or not it was a big change to the sex-ratio paradigm and the Lloyd (1975)'s model is partly a subjective matter. It was yet the first demonstration that two factors responsible for the structure of mating patterns had to be considered altogether in plant mating evolution. In other words, at least in the Oleaceae, the evolution of sexes and self-incompatibility have to be considered jointly in order to understand the mating system evolution. Previous works speculated this joint evolution, but in a different context, and without any direct demonstration: Ehlers and Schierup (2008) showed that a correlation between gynodioecy and SI could be possible (a theoretical prediction partly supported by data), and that heteromorphic SI (i.e. heterostyly) could evolve to dioecy, but without any direct evidence (Barrett 2019a).

Second, distortion segregation and biased sex ratio at birth was needed to fully explain the near balanced sex-ratio 1:1 in natural populations. In the case of gynodioecy, it was suspected very early that genetic conflicts within individuals, between nuclear and cytoplasmic genomes, could result in the evolution, maintenance and high frequency of females (Lewis 1941). It was further confirmed and demonstrated in many different species (Dufay and Billard 2011). In *P. angustifolia*, for the first time, it was demonstrated that genetic conflicts among nuclear genes resulted in the maintenance of high males frequency in an androdioecious species. This changes the nature of models that could be used to study such a situation. Indeed, Lloyd (1975)'s model is phenotypic, which means that the genetic architecture underlying the sex determination can be neglicted, but only under the hypothesis that there are no such evolutionary conflicts at the level of the genes themselves. In this case, taking explicitly into account the genetic architecture is needed, as did Billiard et al. (2015) by modifying Husse et al. (2013)'s population genetics model. But here again, whether or not this should be considered as a big change in the paradigm is a matter of taste. Yet, it means that a whole category of models, phenotypic models, cannot be used, at least in the case of *P. angustifolia*.

Aside from the two previously exposed changes made to the paradigm to solve the *P. angustifolia* paradox, many new questions were raised. First, what are the consequences of the necessity to jointly consider the evolution of SI and sexes? To what extent can it be extrapolated to other plant families? How does it work at the genomic, physiological and morphological levels? Second, even though DSI was discovered in all the other Oleaceae species that were checked so far, all species had their own surprising specificities. *Fraxinus ornus* is androdioecious in some populations but cryptically dioecious in others because one of the two SI groups was lost (Vernet et al. 2016). *F. ornus* populations also show high male frequencies, as high as in *P. angustifolia*, but there is so far no evidence of segregation distortion in this species. *Ligustrum vulgare* shows self-compatible hermaphrodites, but in a single direction: self-compatible hermaphrodites belong to the $G_a$ SI group, they can self-fertilize, they can sire $G_a$ and $G_b$ SI hermaphrodites, can be sired by $G_b$ SI hermaphrodites, but can not be sired by $G_a$ SI hermaphrodites (De Cauwer et al. 2021). *Fraxinus excelsior* populations show a quantitative and continuous variation of the allocation to male and/or female reproductive organs, but structured into two SI groups: one group tends to contain hermaphrodites producing a large number of male flowers and a few hermaphroditic flowers, while the other group contains hermaphrodites producing a majority of female or hermaphroditic flowers (Saumitou-Laprade et al. 2018). *Olea europeae* shows only hermaphroditic individuals, with no unisexuals and no self-compatible phenotype (Saumitou-Laprade et al. 2017a; Besnard et al. 2020; Mariotti et al. 2021), a situation which is expected to be theoretically unstable (Van de Paer et al. 2015). One can speculate that the existence of DSI opens evolutionary pathways that were not anticipated until now. In particular, it is not known so far which one of the situations encountered in the Oleaceae species are evolutionary stable or on their way to dioecy, or other mating strategies.

Finally, maybe the most challenging questions are: How did the DSI evolve first, and how is it maintained? Since the species at the roots of the Oleaceae phylogeny are heterostylous, and that all homostylous species derived after a polyploidisation event, one can speculate that DSI is heterostyly's 'ghost'. However it remains to be demonstrated. More problematic is the open question of the conditions of maintenance of only two SI groups. Indeed, SI always shows dozens of SI groups in all other SI angiosperms families. Indeed, a new SI group is necessarily advantaged while it is rare, because it has a higher number of mating opportunities than frequent SI groups, which thus facilitates the emergence of new SI groups. Two SI groups is the most favorable situation where a new SI group can emerge. But, in the case of the Oleaceae, only two SI groups are observed, and this situation is certainly million years old (Vernet et al. 2016). Hence, solving the paradox in *P. angustifolia*, finally opened new theoretical questions for which the actual sex ratio paradigm has no answer. As will be argued in the following on the bases of Kuhn's criteria, these new open questions justify on their own that resolving the *P. angustifolia* led to a paradigm shift and not simple paradigm adjustments.

## II.    Analysis of the *P. angustifolia* controversy according to Kuhn's theory

Our first goal is to address whether or not the controversy about the mating system of *P. angustifolia* was indeed a *paradigm crisis*, in the sense of Kuhn. A paradigm crisis usually emerges from an anomaly. Before specifically analyzing the *P. angustifolia* controversy, we define the central concepts of Kuhn's theory: *paradigm*, *crisis* and *anomaly*.

**Paradigm.** A *paradigm* (later called a disciplinary matrix in *Reflection on my critics* Kuhn, 1970) is a set of theories, rules and tools adopted - often tacitly - and put into practice by a community of scientists. Paradigms show two main features:

1) A greater explicative power than other scientific activities;
2) A proposition of new problems, or new approaches to extant problems.

When a paradigm change occurs, it is important that most of the progress previously acquired is conserved. In the case of the *P. angustifolia* controversy, the new paradigm should keep all the progress achieved by the theories of Fisher (1930), Lloyd (1975) and Charlesworth and Charlesworth (1981). A paradigm shift is thus not a *tabula rasa*.

*Normal science*, also called *paradigm-based research*, is the application of a paradigm where a majority of researchers in a given scientific field attempt to solve particular problems (Figure 2). Normal science and paradigms are exchangeable. Hence, there are as many paradigms as normal science communities, and there are paradigms at all scales, as long as paradigms are required by a scientific community.

Kuhn (1962) exposed canonical paradigms such as Ptolemean cosmology, Newtonian physics, phlogistics, Einstein's relativity or quantum mechanics. These general (or high level) paradigms are keystones to multiple disciplines (Figure 1). Local or small-cale paradigms also exist, such as the one applied to the mating system evolution in *P. angustifolia*. However, paradigms at any scale obey the same mechanisms (Kuhn, 2012; Hacking, 2012). In his postscript, Kuhn (2012) indeed distinguished four levels of scientific communities where paradigms are applied (Figure 1) :

1) All natural sciences, where the most general paradigm exist (*e.g*, units of measurements);

2) General fields, where global paradigms are adopted (*e.g*, Darwinian theory of evolution for Biology);

3) Major subgroups such as ecology or molecular biology, where paradigms include elements of technology or theoretical tools (*e.g*. Lotka-Volterra and Wright-Fisher models, the concept of fitness as a predictor of genetic changes, the genotype-to-phenotype unidirectional path, or the Fisher's sex ratio paradigm);

4) Specific research groups such as plant population biology, where local paradigms are adopted (*e.g*. some species are biological models such as *Arabidopsis thaliana*, or the ovule/pollen ratio as predictors of mating systems in plants, or the sex-ratio predicted in androdioecious plant populations after Lloyd 1975).

The concept of *paradigm* is thus adapted in the case of the androdioecious status of *P.*

*angustifolia*, even though the associated scientific community interested in this question is small. While Fisher's sex ratio theory might be considered as a level 3 paradigm, what we call the Fisher-Lloyd-Charlesworth and Charlesworth paradigm is a level 4 paradigm, since it was built to specifically explain plant mating system evolution (Figure 1). On a side note, this raises the question whether a level 1, 2 or 3 paradigm could be refuted by an experiment which necessarily takes place within a level 4 paradigm. As Duhem (1906) pointed out, the different scales and the interdependence of paradigms challenge the notion of *experimentum crucis* and refutability.

**Anomaly.** Anomaly refers to an object that does not fit into the paradigm and whose resolution is not possible within the paradigm it emerged from. In the evolutionary framework of Kuhnian models, anomalies are analogous to evolutionary pressures in ecology by negatively selecting a theoretical set's particular locus. Concerning *P. angustifolia,* the local paradigm that ruled normal science seemed to have been set since 1930 by Fisher and reinforced later on (Lloyd 1975; Charlesworth and Charlesworth 1981). Therefore, *P. angustifolia* is recognized as an anomaly since it doesn't match the paradigm's prediction (Pannell 2002). This characterisation is amplified by the unusual features of *P. angustifolia*. In general, objects with rare characteristics tend to be naturally opposed to paradigms as the paradigm's goal is to be as general as possible in a particular field. The recognition of an anomaly is chiefly a subjective and collective process, that does not imply that what is identified as an anomaly actually is one (Watkins, 1970; Kuhn, 1970).

**Paradigm crisis.** When an anomaly is identified and collectively accepted as such, a paradigm crisis usually occurs. A paradigm crisis is not a period of downfall for science but instead leads to a proliferation of new propositions and discussions about the foundations of a paradigm. The start of the *P. angustifolia* crisis took place around 1992 (Lepart and Dommée 1992), while it ended around 2015 (Saumitou-Laprade et al, 2010; Pannell and Korbecka, 2010; Billiard et al. 2015).

A paradigm crisis ineluctably leads to what Kuhn stated as *extraordinary science,* in opposition with *normal science*, where scientific activity and production are decreased, and where a scientific field becomes unstable. Extraordinary science naturally causes a division in the formerly solid adhesion to a certain paradigm in a given group of scientists. The number of newly formed groups is very variable and sometimes depends on socio-geographic parameters or local scientific tradition. Sometimes there are as many scientific groups as laboratories dedicated to a discipline facing a crisis. Concerning *P. angustifolia* two research groups were opposing (see Section I for details): the first one, more in agreement with Lloyd's paradigm, assure that based on the validity of the model, *P. angustifolia* had to be cryptically dioecious; The second one, based on laborious and clarifying experimentations, assured that *P. angustifolia* was androdioecious, yet without getting rid of the paradoxical situation and the anomaly.

**Adoption of a new paradigm.** Because it is not a stable state of research, extraordinary science consequently leads to a conservation or a replacement of the formerly shared common paradigm. The conditions for the identification of a paradigm change are (Conant 1947, Kuhn 2012, Wray 2021):

1) The competing paradigm must resolve a primordial problem or an anomaly, *e.g.* the unexpected mating system in *P. angustifolia*;
2) The competing paradigm must conserve the majority of the previously acquired knowledge, *e.g.* the DSI and distortion segregation were included into the Fisher-Lloyd-Charlesworth and Charlesworth's models.

3) The competing paradigm must open new problematic fields, called *puzzles* by Kuhn, that nevertheless do not challenge the validity of the competing paradigm itself, *e.g.* the evolutionary origin and maintenance of the DSI is still puzzling (see below).

4) The competing paradigm must propose ideas contradictory to the former one (Conant 1947).

Criteria 2) and 4) seem contradictory at first sight: how can a paradigm shift conserve knowledge while contradicting it? This particularity is due to the evolutionary approach inherent to Kuhn's model. Anomalies are indeed submitted to a pressure analogous to natural selection in biological evolution, which plays on a particular locus of a genome. Anomalies can be considered as a particular locus of a paradigm which suffers epistemological pressure while leaving the rest of the paradigm relatively untouched. Criteria 2) and 4) are therefore compatible since what is contradicted and negatively selected in the old paradigm is only what the new paradigm resolves, while the main corpus of knowledge remains unthreatened.

**Incommensurability**. The competition between the former established paradigm and the propositions of a new paradigm is actually more complex due to what Kuhn called the incommensurability of paradigms (a central concept in Kuhn's theory strongly debated in philosophy of science, Shapere, 1944 ; Ruse, 1989 ; Lakatos 1976 ; Watkins 1970). The general idea of incommensurability is that each scientific discourse is settled within a particular theoretical framework that changes particular dimensions of the world. Four different types of incommensurabilities can be distinguished: (Kuhn 1962 & 2000, Wray 2021, Devlin 2021):

1) Ontological and lexical (see also Ruse 1980): paradigms create and carry a set of conceptual beings, while other types of beings can not be conceptualised. In the *P.*

*angustifolia* controversy, the same being could be at the same time dioecious for a given paradigm and androdioecious for the other. This dimension affects the way scientists see things, and the relationship they establish between these things;

2) Methodological: when beings differ, because of ontological incommensurability, the tools and methods to study them are also different. In the case of *P. angustifolia*, paternity analysis, diallelic crosses, or explicit genetic modeling were necessary to resolve the paradox in addition to Lloyd (1975)'s phenotypic models;

3) The incommensurability between specialty communities: the communication and the research on a common ground are nearly impossible since the theoretical and functional frameworks of research groups are too distinct from one another. This was especially important when the controversy about the mating system in *P. angustifolia* was extended to the whole Oleaceae family, and especially in the case of the olive oil (*Olea europaea*) where one group was mostly constituted of theoreticians in evolutionary biology while the other groups came from agronomy and plant physiology academic backgrounds;

4) Dissociation: It occurs when writings from an earlier era are not comprehensible under our modern gaze. As the *P. angustifolia* controversy is recent and temporally narrow, this dimension of incommensurability is not relevant.

The problematic aspect of incommensurability comes from the idea that two scientists in two different theoretical frameworks live in two distinct worlds while still looking at the same reality. At best, the different frameworks partially overlap, but they can be mutually exclusive. However, incommensurability should not be confused with incommunicability. Two scientists in distinct theoretical frameworks can discuss together, but it requires a translation from one lexicon to the other. It even requires a transformation since both of the two participants have to try to see the world as the other sees it (Kuhn, 1970, p. 277). This transformation is made possible in Kuhn's theory by its idealistic framework (see the "model *vs*. data" section below for details and a discussion of that point in the context of the *P. angustifolia* controversy). On a side note, translation and transformation make *Gestalt* switch possible for philosophers, thus allowing them to roam and temporarily settle within two concurrent paradigms.

Overall, the paradigm change which occurred in *P. angustifolia* can be compared with the discovery of Oxygen at the end of the XVIII[th] century (Kuhn 1970). In 1774, Joseph Priestley identified the gas collected from heating red oxide of mercury as dephlogisticated common air. In 1775, Lavoisier identified the same gas as unaltered air and concluded that it was a constituent of the atmosphere. Perceptions and interpretations of the same object by Priestley and Lavoisier were incommensurable: they were two different paradigms and two distinct scientific "worlds" (see also Ruse 1989). A similar situation occurred in the case of *P. angustifolia,* based on the opposition of different perceptions and interpretations of the same observations: one research group considered *P. angustifolia* was a cryptically dioecious species *because* it showed a 1:1 sex ratio; the other considered *P. angustifolia* was androdioecious *because* hermaphrodites had a siring success similar to that of males.

**Internal resilience and external resistance.** The incommensurability of paradigms prevents from adopting a new paradigm solely based on rational arguments and mathematical reasoning. The stability of a paradigm actually depends on two variables:
1) Internal resilience (introduced by us as a synthesis of Kuhn's main ideas): *e.g.* the paradigm general coherence, its ability to easily incorporate new phenomena, the amount of theories whose existence depend on it since all paradigms are entangled vertically and horizontally, etc. (Figure 1);

2) External resistance to the crisis: *e.g.* to what degree researchers are attached to the paradigm, its place in the learning process of new students, its historical background, its prestige, etc.

External resistance mechanisms inevitably lead to a form of 'scientific crystallization'. When a set of theories are clearly adopted as paradigmatic, they are immediately endowed with a high resistance to anomalous data. Such resistance have multiple possible external sources, since the theoretical core of the paradigms are not drastically changed in the process of adoption. Hoquet (2020) for example showed that resistance in favor of Bateman's principle was mostly due an excessive amount of citations of Bateman (1948)'s seminal paper. Scientific crystallization is not to be understood as a pejorative term in this context: it is the tendency to give more confidence and value to the paradigm in times of crisis than in times of normal science. Even though crystallization might lead to regrettable outcomes (*e.g.* the fervent adhesion to a recently abandoned paradigm), it is a vital mechanism to conserve the stability of normal science. Even though scientific crystallization has probably been encountered in the case of *P. angustifolia*, the new paradigm (*i.e.* the existence of the diallelic self-incompatibility and the distortion segregation bias in *P. angustifolia*) is now largely adopted by the community (Pannell and Korbecka 2010; Barrett 2019b), even though some resistance is still encountered (*e.g.* Breton et al. 2021, see below).

The resistances encountered by the new paradigm in the *P. angustifolia* controversy were certainly mainly due to an application of the principle of parsimony to the designated models, where the costs of hypotheses and mechanisms involved in an explanation are evaluated and balanced. Three elements were differently evaluated by the two competing research groups: changing *vs*. maintaining the Fisher-Lloyd-Charlesworth and Charlesworth's models and predictions; accepting *vs*. rejecting the observations that hermaphrodites were functionally males and females in *P. angustifolia*; introducing self-incompatibility *vs*. ovule discounting as a new mechanism to reconcile high male frequencies in natural populations and models' predictions.

The two competing research groups reached two opposite conclusions:

1) Either the Fisher-Lloyd-Charlesworth and Charlesworth's models are too costly to be dismissed and high male frequencies in natural populations of *P. angustifolia* are likely explained by imprecisions or misinterpretations of the experimental observations. As a consequence, according to the established paradigm, hermaphrodites in the populations of *P. angustifolia* are females in a cryptically dioecious system. Ovule discounting was hypothesized as a heavy mechanism to reconcile data and predictions;

2) Or rejecting observations is too costly and high male frequencies in populations of *P. angustifolia* are likely explained by the incompleteness of the paradigm. As a consequence, hermaphrodites are functional and the paradigm has to be changed. According to the observations, heavy adjustments (*i.e.* the DSI and genetic conflicts with the distortion segregation bias) had to be included in the established paradigm.

Even though a heavy new mechanism had to be considered in both propositions (ovule discounting *vs*. DSI and distortion segregation), the principle of parsimony indicates that proposition 2) was heavier than proposition 1) because of the cost of rejecting the established paradigm in proposition 2).

According to Kuhn's theory, it is always possible to challenge the status and the validity of an observed anomaly. It is only when the anomaly is recurrent in time and in different experimental protocols that models' parsimony have to be questioned, and that more costly mechanisms have to be considered. In the case of *P. angustifolia*, it appears that the anomaly and the paradigm crisis needed paradigm changes by the costly introduction into the paradigm of new surprising mechanisms: the DSI (an unexpected homomorphic SI system) and distortion segregation (unexpected conflicts between *nuclear* genes involved in sex and SI determination).

**III. The present and future status of the paradigm change in plant mating system evolution**

**Is the discovery of DSI in *P. angustifolia* a paradigm shift or a paradigm adjustment?**

The identification of a paradigm shift requires four *sine qua non* conditions (Kuhn 1962; Wray 2021): i) The conservation of the previously acquired knowledge; ii) The resolution of an anomaly; iii) The opening of a new problematic field, since new paradigms are always formed around a set of problems; iv) the new paradigm partly contradicts the former one.

We propose that a paradigm is only *adjusted* if only conditions i) and ii) are fulfilled, but not conditions iii) and iv). Hence, conditions iii) and iv) are necessary and sufficient for shifting a paradigm (Table 1) since the contradiction of the former paradigm consequently opens an area of research that was not revealed before. The choice of these two necessary and sufficient criteria can be discussed but nonetheless offers a framework under which paradigm dynamics can be subsumed while being directly extracted from Kuhn's and Conant's works. As Shapere (1964) stated it: "*where do we draw the line between different paradigms and different articulations of the same paradigm?*". The problem of the identification of necessary and sufficient criteria to characterize a paradigm shift has always been present in Kuhn's theory. Nonetheless, it does not mean that it is impossible to find efficient criteria to operate an analysis.

Regarding *P. angustifolia*, condition i) is fulfilled because the seminal paradigms from the sex ratio theory and plant mating system evolution by Fisher-Lloyd-Charlesworth and Charlesworth have been conserved; Condition ii) is also fulfilled because the anomaly that initiated the crisis about the status of the mating system of *P. angustifolia* has been solved. However, the discovery of the DSI in *P. angustifolia* was not the end of the story since the DSI became an object of study on its own. Indeed, the DSI generated new problems not yet explained by the new paradigm (the evolutionary origins and maintenance of the DSI in the whole Oleaceae family and beyond, and the possible consequence that the evolution of sexes, self-incompatibility and genetic conflicts are generally linked, Billiard et al. 2011; Barrett 2019a, 2019b). Hence, condition iii) is also fulfilled. Condition iv) is also fulfilled since we showed that the two propositions "*P. angustifolia* is dioecious" and "*P. angustifolia* is androdioecious" are contradictory. Finally, we conclude that the discovery of DSI in *P. angustifolia* should be considered as a paradigm shift, and not only as a paradigm adjustment, contrary to Pannell and Korbecka (2010)'s analysis.

Moreover, in his 1970 contribution, Kuhn mentions another criterion that might be called the *consequential* criterion: *"The gist of the problem is that to answer the question 'normal or revolutionary?' one must first ask, 'for whom?''*. (Kuhn, 1970, p. 252) This criterion tries to take account of the *locality* of shifts. It states that to call something revolutionary — *e.g.* to identify a paradigm shift — we have to consider how the practice of science has been affected relatively to each scientific group. In the case of *P. angustifolia*, while the discovery of DSI did not change anything to the way most biologists of evolution practice science, it drastically changed the way groups dedicated to plant mating systems pursued their research. As Kuhn stated it: *"Many episodes will then be revolutionary for no communities, many others for only a single small group, still others for several communities together, a few for all of science"* (Kuhn, 1970, p. 253).

**A variety of resistances to the paradigm shift.** Resistance to changes is inherent to paradigms as Kuhn's theory suggests. We presented one such resistance from one research program, which was resolved between 2010 and 2015. Our demonstration focused on the opposition we thought was the most important, *i.e.* cryptic dioecy *vs.* androdioecy. However, other resistances to the paradigm can be identified. We present here three categories of opposition, playing at different times. First, a resistance which appeared instantaneously after the paradigm shift but which has known no posterior developments. Second, a resistance which runs over several years and is still ongoing because one research program only partly accepts the DSI paradigm shift in the Olive tree. Third, since new resistances necessarily emerge once a paradigm shift is adopted, we propose one such possible new resistance that can slow down future discoveries in the Oleaceae.

*Cryptic distyly.* A new resistance instantaneously appeared after the DSI was adopted. Pannell and Korbecka (2010) concluded that the discovery of the DSI was indeed a paradigm change but, at the same time, that "*P. angustifolia* is not cryptically dioecious, but displays something akin to cryptic 'distyly'". Their conclusion was however ambiguous. It would mean that nothing was new under the sun as distyly is common in angiosperms. In other words, if *P. angustifolia* was indeed cryptically distylous, then the discovery of the DSI is finally not a paradigm change. Calling for cryptic distyly can thus be interpreted as a resistance to the paradigm change. However, resisting the paradigm change with cryptic distyly raises unnecessary difficulties. Distyly is by definition heteromorphic: heteromorphic populations show individuals with either long or short styles that coevolved with their animal pollinators. How can heteromorphism be cryptic? What about Oleaceae species not pollinated by animals? Most importantly, adopting cryptic distyly in *P. angustifolia* would have oriented future research about the evolution of the mating system in Oleaceae to a single question: how heteromorphic distyly has been lost while SI is conserved? On the contrary, adopting the DSI makes the latter question only a possible evolutionary scenario among others, which are still to be investigated. Adopting the DSI paradigm also raises a general question about the relationships between homomorphic self-incompatibility, heterostyly and sexes (Barrett 2019a,b).

*Self-incompatibility in the olive tree.* Determining the mating system of the olive tree (*Olea europaea*) is a long-standing issue which involved several communities of researchers (*e.g.* Wu et al. 2002; Breton and Bervillé 2012; Saumitou-Laprade et al. 2017a). Observations regarding the mating system in *O. europaea* appeared complex and confusing: the species seemed at the same time self-incompatible, with some cultivars capable of recurrent self-fertilization under some crossing conditions, while other cultivars were only partially self-incompatible. This led a research group to propose a complex organization of the SI system with dozens *S*-alleles, expressed at the sporophytic level, with complex dominance relationships between *S*-alleles (Breton et al. 2014, 2016). The discovery of the DSI dismissed such a complex SI system (Saumitou et al. 2017a, 2017b; Besnard et al. 2020; Mariotti et al. 2021). However, as predicted by Kuhn's theory, there was, and there is still, an ongoing resistance against the DSI paradigm as a relevant mating system in the olive tree (Breton et al. 2017; Farinelli et al. 2018; Breton et al. 2021).

The still ongoing resistance about the DSI in the Olive tree can be explained by two different mechanisms. First, because of sociological reasons. The concurrent research program is led by scholars involved in the study of the olive tree for dozens of years, especially for industrial and agricultural purposes. Second, for some scholars, the mating patterns observed in the Olive tree are still unexplained by the DSI (Farinelli et al. 2018; Breton et al. 2021). These authors concluded that the crossing patterns observed in the olive tree are, on the one hand, too complex to be compatible with the DSI and, on the other hand, DSI can not explain that self-fertilization is recurrently reported in some cultivars. However, it has been shown that self-compatible genotypes can stably coexist with the DSI, both theoretically (Van de Paer et al. 2015) and empirically in *Ligustrum vulgare* populations (De Cauwer et al. 2021), which makes the existence of self-compatibility in the olive tree perfectly compatible with the DSI paradigm. Yet, the situation in the olive tree is certainly different than in *Ligustrum vulgare*: a self-compatible genotype has been identified in the latter, while self-compatibility seems to be not associated to a particular genotype in the former, suggesting that self-compatibility might be a plastic response in the olive tree.

The disagreement between the DSI paradigm's predictions and mating patterns reported in the olive tree can be explained by observation errors in experiments, with at least two different origins. The cultivars used in the controlled crossing experiments could be composed by more than a single genotype as putatively supposed (Mariotti et al. 2021). If a cultivar indeed contains a single genotype, crosses between different individuals from the same cultivar would not be possible under the DSI. Hence, a compatible cross within a cultivar might indeed be an evidence against the DSI. However, this conclusion can be reached only if the genetic composition of the cultivar is verified and controlled, which needs a particular experimental design. Without such a perfectly controlled experimental design, where cultivars are indubiously associated to a single genotype, crosses between and within cultivars can be found compatible or incompatible because apples would be compared to oranges (Bergelson et al. 2016; Saumitou-Laprade *et al*. 2017b).

Another possible origin of experimental errors could come from crosses not controlled enough. *O. europaea* is wind-pollinated, which means that the air surrounding trees can be saturated in pollen, especially by long-distance pollen. If inflorescences are not bagged early enough, flowers might have received pollen from many different trees, especially by compatible trees, which can lead experimenters to false interpretations. Overall, the adoption of the new DSI paradigm leads to casting doubts on the validity of empirical observations, and legitimates requests to verify the robustness of the data reports (*e.g.*

validation of crosses by paternity analysis, Saumitou-Laprade *et al*. 2017a, 2017 b; Mariotti et al 2021). Ironically, the validity of the observations was one of the arguments against the interpretation of *P. angustifolia* as an androdioecious species. This shows that even though paradigms change, the nature of resistances do not.

*New paradigm, new resistance: a neverending rise of experimental standards?* Once a new paradigm is adopted, scientific activities are reorganized and science returns to its normal state (Figure 2), which involves new resistances to paradigm changes. Kuhn's theory can thus not only help in understanding a paradigm change, but also help in anticipating and identifying new established resistances. The ongoing debate about the mating system in *O. europaea* highlights such an issue.

The large adoption of the DSI paradigm is accompanied in particular by the prediction that by default an Oleaceae species should show two SI groups, possibly in association with an additional mating phenotype such as males, as in *P. angustifolia*, or self-compatible individuals, as in *L. vulgare*. Yet, the existence of the DSI has been demonstrated in a few species only. It is always possible, in principle, to find a species with a different mating system, for example with more than two SI groups. Since DSI is now the default situation, a research program that would aim to demonstrate, for example, the existence of at least a third SI group of hermaphrodites, would be faced with a high resistance: only high standard experimental results would convince the new paradigm defenders that DSI is not general as actually thought. What are these high standard experiments? The most important requirements would certainly be high standard controlled crosses: on a sufficiently large number of different individuals, with all needed treatments (two positive controls showing that the stigma of the pollen receiver and the pollen of the pollen donor are functional, and two negative controls showing that self-pollen or crossing between individuals of a given group give no fertilization), such that crosses are indubiously controlled enough (pollen contamination should be excluded), definitely assessed by paternity analyses.

Such high standards experimental conditions have been progressively built to defend the DSI as a factor explaining the situation encountered in *P. angustifolia*. Now that the new paradigm has been adopted, these experimental conditions have become the new experimental standards. We can thus speculate that demonstrating an exception to the DSI would need to push even further these experimental standards in order to convince the DSI defenders.

**Models vs. data.** The predominant role of models over data in the discussions around *P. angustifolia* is striking. To a certain degree, Fisher (1930)'s sex ratio paradigm was the main element of verification since data were faced with its predictions since it led one research group to cast doubts on the validity of observations. Why were models so predominant over data in the case of *P. angustifolia*? Despite such a predominance, why the paradigm shift could occur against models?

Experiments are often naively considered as testing a theory's predictions. However, in the *P. angustifolia* controversy, the reverse occurred since experiments were considered dubious because not in agreement with theoretical predictions: the predictions from the Fisher-Lloyd-Charlesworth and Charlesworth's models were used to challenge data interpretations. This turnaround could be due to the nature of experimentation in macro-biology. In physics, for example, the canonical method is relatively adequate since it is easier to perform controlled experiments with isolated objects. In macro-biology, experimental conditions are less easily controlled. In particular, performing experiments of macro-biological close and isolated systems is in practice almost impossible. Models and data are in a balance where the decrease of trust in the validity of experimental conditions increases the trust in models. This raises the general issue about the relationship and agreement between models and data: How can we know whether or not ingredients of models are coherent with observations from the physical world?

The concept of "world" proposed by Kuhn partly addresses this question. Each paradigm adopted by a scientific group corresponds to different scientific worlds, where what is observed is directly influenced by its own set of theories (*e.g.* Priestley and Lavoisier about oxygen). However, different *worlds* are not to be conflated with different *interpretations* (see Devlin 2021 and Fodor 1984, 1988 for the distinction between *weak* and *strong* theory-dependence). Lavoisier did not consider the gas extracted from heating red mercury oxide *as* pure air, it *was* air. In the case of *P. angustifolia,* one group *saw* an androdioecious species, the other one *saw* a cryptically dioecious species. The paradigm thus provides an *a priori* signification to data themselves, which can consequently explain the predominant role of models over experiments (Figure 3).

This incommensurability concerning perception and interpretation is clearly stated by Kuhn (1970, p. 276). The ontological baggage that paradigms carry is inseparably linked to Kuhn's theory of language. One of the main concerns of Kuhn's work is indeed to determine the specificity of the scientific languages, and how the terms developed by these *attach to nature*. The idea of *world* in Kuhn's work consequently contains a theory of reference, that permits to establish a relation between a supposed external world, and a *lexicon* that we attach to it. This point is illustrated in the "double coinage" metaphor (Kuhn, 2000, p. 29). Actually, the familiarity between Kuhn and idealist philosophy is something that has already been underlined by Ruse (2008) and Wray (2021) — see for example Kuhn's referring to Kant's *ding an sich* in 1991.

All the above leads to questioning the very nature of *data*, in particular in the case of population and evolutionary biology, as illustrated here by the *P. angustifolia* controversy. A long tradition of philosophers, starting from Kant and his *Critique Of Pure Reason* (1787)*,* followed by Schopenhauer (1813, 1844), Duhem (1906), Bachelard (1938), Kuhn (1970) and Hacking (2000), among others, suggested that the notion of "raw data" was dubious. Supposing that knowledge can be built from raw data means that it is possible to extract pure elements of intelligible reality that can be included in theoretical models afterwards. Instead, a less radical approach to experimentation seems to be more adequate: models pre-determine experimentations and data are elements of theories that we inject into the empirical reality. This has two implications. First, the interpretation of data varies according to the paradigm; Second, the very nature of data itself is also determined by the paradigm (Kuhn 2012, Ruse 2008). This can explain the origin of the incommensurability of the paradigms in the case of *P. angustifolia* which led to the paradigm crisis.

However, this does not lead to a total relativism where "everyone has its own equally viable view of the world". Scientific discourse is mainly descriptive, and description is principally a matter of abstract properties applied to concrete or abstract objects. Proposing that scientific change is not only a matter of logical reasoning does not mean that there is no argument to persuade a colleague to convert to another paradigm (Kuhn 2012). As previously mentioned, internal criteria can be found, such as the amount of theories that rely on the existence of a given paradigm This internal dimension is also why Kuhn should not be misidentified as a relativist but rather as a functionalist. His concern was *in fine* to determine *why* does science works, and fruitfulness — a criteria that Conant (1947) also proposed — is maybe the internal golden standard of Kuhn's theory.

The notion of world also raises a concern that comes from kuhnian models. If paradigms are two distinct worlds where scientists evolve, then conversions from one paradigm to another should not be possible, because it would break the perceptual frameworks in which these individuals have been settled. In most of the examples used by Kuhn, namely the controversy between Lavoisier and Priestley, the tenants of the former paradigm stayed attached to it until they died. This leads to the idea that conversion between paradigms should not be possible, reinforcing the idea that a paradigm change really occurs once the tenants of the former one die. However, in the case of *P. angustifolia,* we can observe that these conversions have occured, some tenants of the former paradigm have "changed their worlds" to the new one. Is this a refutation of the notion of incommensurability, or a refutation of the identification of the DSI as a change of paradigm ? Neither of them.

We firstly have to mention that Kuhn was perfectly aware of the fact that conversions can occur:

*Note, however, that the possibility of translation does not make the term 'conversion' inappropriate. In the absence of a neutral language, the choice of a new theory is a decision to adopt a different native language and to deploy it in a correspondingly different world.* (Kuhn, 1970, p. 277)

However, Kuhn did not specify that conversion is chiefly an effect of scale. Since each new paradigm imposes a distinct world from the previous one, and that some paradigms are broader than others (Figure 1) we can suppose that conversions are seldom or absent at level 2, while being much more common at levels 3 and 4. Indeed, the broader a paradigm is, the more radically distinct the world it proposes is from the one it replaces, making conversions impossible for the mind of a single scientist. In lower scales however, the change of a paradigm only affects particular *loci* of reality. This finding is notably one of the main reasons why the Kuhnian approach should be more often applied to low-level paradigms.

**Unexpected origins and consequences of paradigm shifts.** The paradigm shift generated by the resolution of the *P. angustifolia* controversy also illustrates how the scientific analysis of an object with high heuristic assets can have unpredictable economic impacts. This shows that the putative economic value of an object is certainly not a sufficient *criterion* to decide its scientific value. The economic value of *P. angustifolia* is not intrinsic, but especially depends on its phylogenetic relationship with the Olive tree. As many other species or specific problems could be studied in close relationships with the Olive tree, the economic value of *P. angustifolia* could not be predicted. Even though the origin of the paradigm change was certainly independent of any economic considerations in the case of *P. angustifolia*, this case-study suggests that paradigm changes can lead to discoveries with potentially important economic consequences. The demonstration of the DSI in the olive trees (Saumitou et al. 2017a), as expected under the new paradigm, initiated a reflection about how the olive trees orchards should be organized, composed and managed in order to improve and better control the quantity and quality of the production of olive fruits.

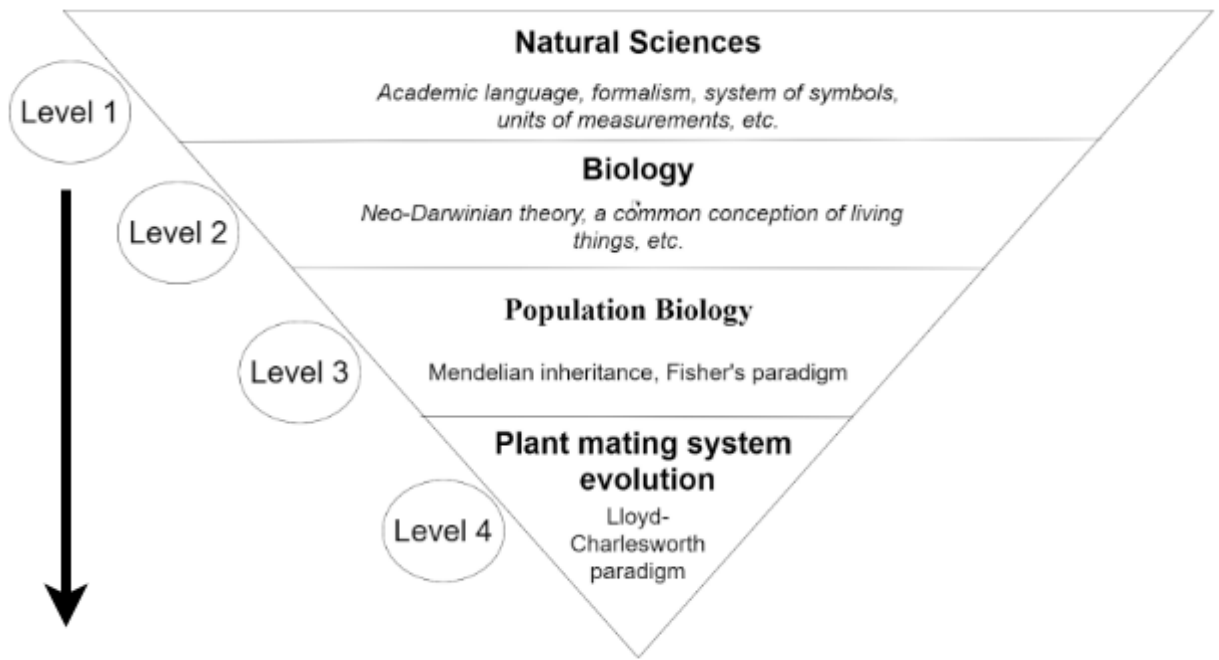

Figure 1: The entanglement of scientific fields and their shared paradigms applied to plant mating system evolution. This schematic representation shows that the field of plant mating system evolution and its paradigms (Level 4 paradigms) are nested within Population biology (Level 3 paradigms), which is nested within Biology (Level 2 paradigms) and Natural sciences (Level 1). Note that the representation is not exhaustive as upper or lower levels could be added.

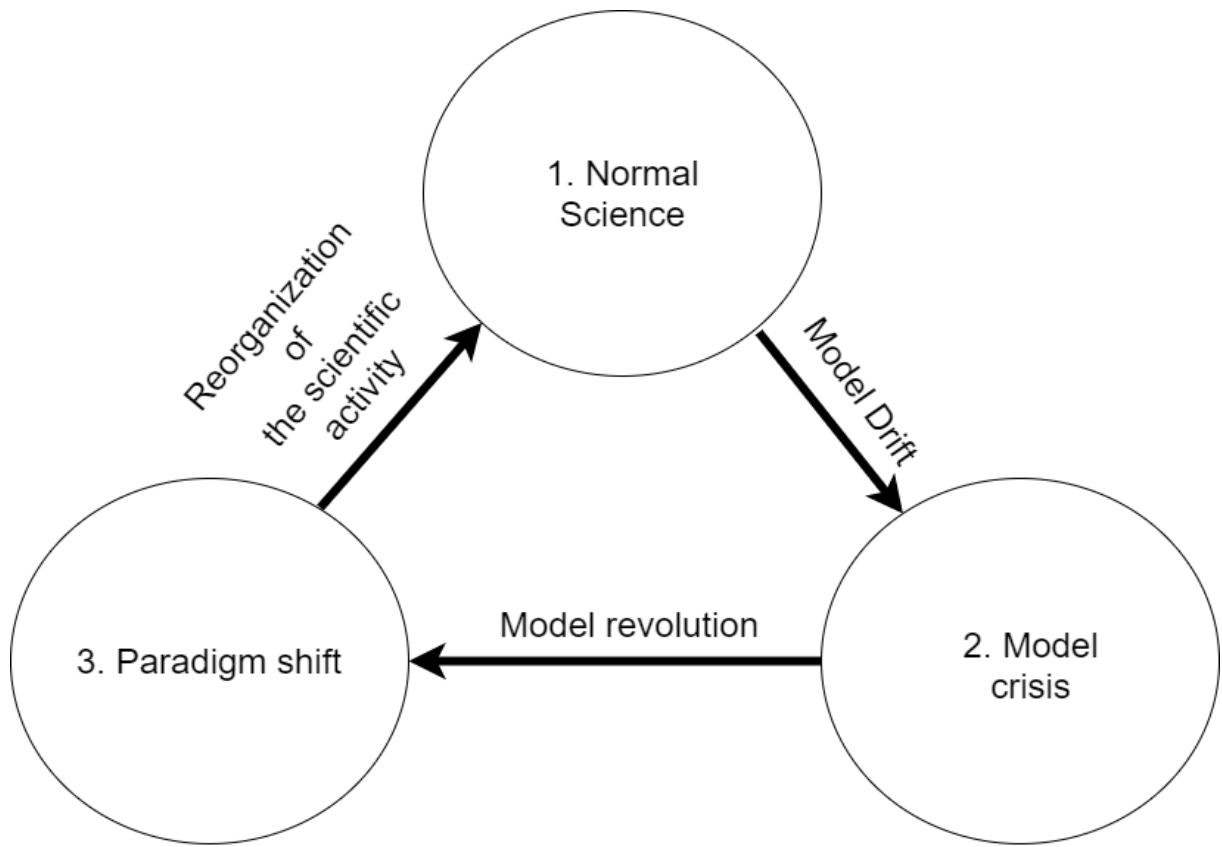

Figure 2 : The Kuhn cycle of scientific activity. **1.** *Normal science*: the sex ratio and plant mating system evolution paradigm according to predictions by the Fisher-Lloyd-Charlesworth and Charlesworth paradigm. **2.** *Model crisis*: the unexpected and unexplained high frequency of males in P. angustifolia populations was an anomaly, which led to a paradigm crisis with two opposing perceptions and interpretation of data: *P. angustifolia* was either indeed androdioecious, *or* it was cryptically dioecious in agreement with the established Fisher-Lloyd-Charlesworth and Charlesworth paradigm. **3.** *Paradigm shift*: The combination of experimental results and new models led to the acceptance that two new mechanisms should be considered to explain the high male frequencies in *P. angustifolia*: the Diallelic Self-Incompatibility system and sexual distortion segregation. These discoveries were followed by new theoretical and empirical questions regarding the origin, maintenance and coevolution of the two mechanisms, and the mating system status in the other Oleaceae species.

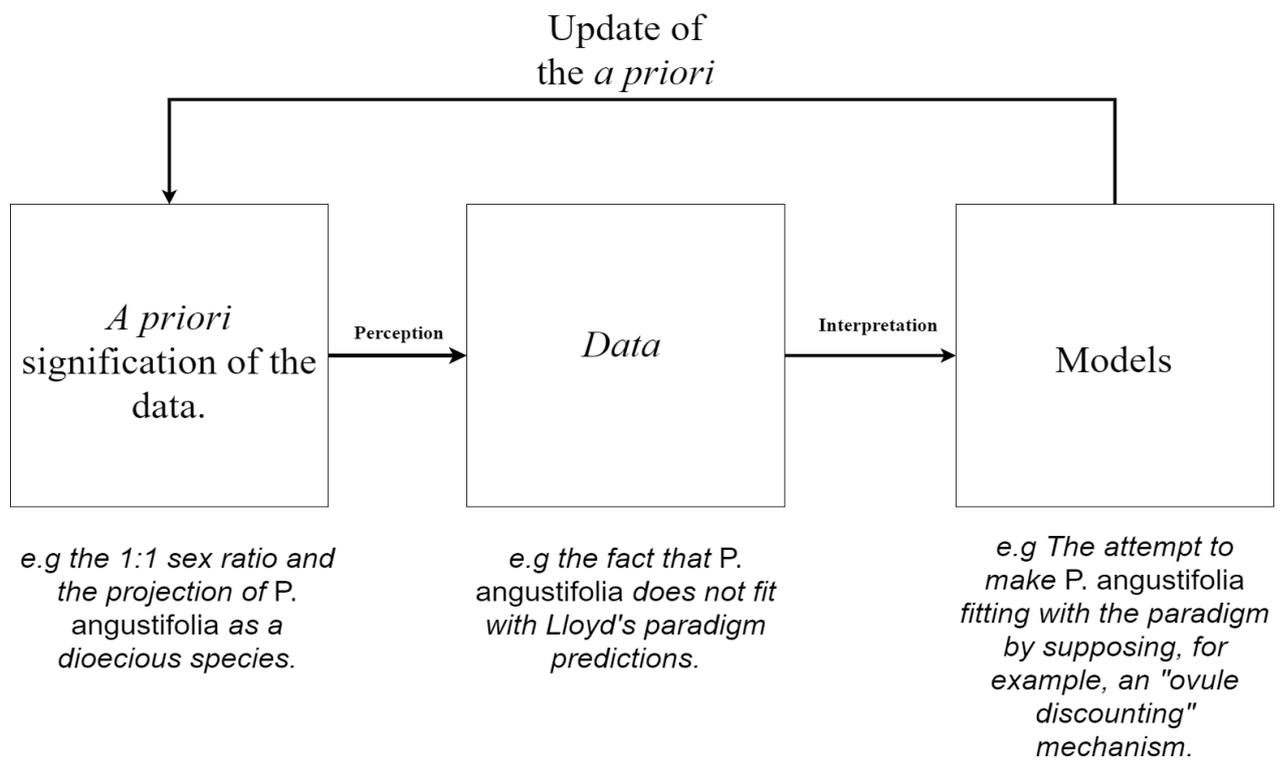

Figure 3: Interactions between data and models.

| | **Paradigm adjustment** | **Paradigm shift** |
|---|---|---|
| **Conservation** of the previously acquired knowledge | ✓ | ✓ |
| **Resolution** of an anomaly or of a fundamental problem | ✓ | ✓ |
| **Opening** of a new problematic field | X | ✓ |
| Conant's criteria: the new paradigm partly or totally **contradicts** the former one | X | ✓ |

Table 1: Proposition of necessary and sufficient criteria to identify a paradigm shift over a paradigm adjustment (after Kuhn 1962, 2012 and Wray 2021).

**Acknowledgments.**

We thank Lucien Platon, for his useful advice and the continual review of our work and Cyprien Cocquyt for his precious reading advice and guidance through the preparation process of this article. We also thank Stefaan Blancke, Koen Tanghe, Pierre Gérard and Gauvain Leconte-Chevillard for precious discussions and advice. We warmly thank Thibault Masset, Jacques Capelle and l'Atelier Critique without whom this work would not have existed. S.B. does not thank the administration team of the University of Lille in charge of the management of students' internships with whom this work would not have existed.